\documentclass{amsart}
\usepackage[dvips]{color}
\usepackage{lmodern}
\usepackage[T1]{fontenc}
\usepackage{stmaryrd}
\usepackage{amsmath}
\usepackage[foot]{amsaddr}
\definecolor{ddgreen}{rgb}{0.2,0.9,0.0}
\definecolor{turco}{rgb}{0,.5,.5}
\definecolor{reddish}{rgb}{.8,.15,.2}
\definecolor{barney}{rgb}{.6,0.12,0.9}
\definecolor{magic}{cmyk}{0,.7,0,.5}
\definecolor{blueish}{rgb}{.1,.1,.8}
\definecolor{greenish}{rgb}{.2,.4,.2}
\begin{document}

\title[Weyl anomaly of GJMS]
{Holographic Weyl anomaly for GJMS operators: one Laplacian to rule them all}
\author{F. Bugini $^{\S}$ and D.E. Diaz $^{\dag}$}
\address{${\S} $ Departamento de Matemática y Física Aplicadas, Universidad Católica de la Santísima Concepción, Alonso de Ribera 2850, Concepción, Chile}
\address{${\dag} $ Departamento de Ciencias Fisicas, Universidad Andres Bello, Autopista Concepcion-Talcahuano 7100, Talcahuano, Chile}
\email{ \hspace*{0.0cm}${\S}$ fbugini@ucsc.cl , ${\dag}$ danilodiaz@unab.cl }
\begin{abstract}
The holographic Weyl anomaly for GJMS operators (or conformal powers of the Laplacian) are obtained in four and six dimensions. In the context of AdS/CFT correspondence, free conformal scalars with higher-derivative kinetic operators are induced by an ordinary second-derivative massive bulk scalar. At one-loop quantum level, the duality dictionary for partition functions entails an equality between the functional determinants of the corresponding kinetic operators and, in particular, it provides a holographic route to their Weyl anomalies. The heat kernel of a single bulk massive scalar field encodes the Weyl anomaly (type-A and type-B) coefficients for the whole tower of GJMS operators whenever they exist, as in the case of Einstein manifolds where they factorize into product of Laplacians.\\
While a holographic derivation of the type-A Weyl anomaly was already worked out some years back,
in this note we compute holographically (for the first time to the best of our knowledge) the type-B Weyl anomaly for the whole family of GJMS operators in four and six dimensions. There are two key ingredients that enable this novel holographic derivation that would be quite a daunting task otherwise: (i) a simple prescription for obtaining the holographic Weyl anomaly for higher-curvature gravities, previously found by the authors, that allows to read off directly the anomaly coefficients from the bulk action; and (ii) an implied WKB-exactness, after resummation, of the heat kernel for the massive scalar on a Poincar\'e-Einstein bulk metric with an Einstein metric on its conformal infinity.\\
The holographically computed Weyl anomaly coefficients are explicitly verified on the boundary by exploiting the factorization of GJMS operators on Einstein manifolds and working out the relevant heat kernel coefficient.\\
\end{abstract}
%\pacs{}

%\submitto{}

\maketitle
\qquad\\

%------------------------------------------------------------------------
\section{Introduction}
\qquad\\

Conformal powers of the Laplacian $P_{2k}$ (or GJMS operators for short~\cite{GJMS92}) are higher-derivative generalizations of the conformal Laplacian or Yamabe operator of the form
\begin{equation}
P_{2k}=\Delta^k+LOT
\end{equation}
with principal part given by an integer power of the Laplacian and complemented by lower order (in derivative) terms (LOT) built up out of the Ricci tensor and covariant derivatives. They first arose within the general Fefferman-Graham program \cite{FG85} induced by the $k$-th power of the ambient Laplacian $\tilde{\Delta}^k$ and allowed Branson's characterization of the Q-curvature in general even dimensions as given by their zeroth order term\footnote{For recent results on recursive relations and explicit construction of GJMS operators and the associated Q-curvatures, we refer to the works~\cite{Juhl11, FG13} and references therein.}~\cite{Bra93,Bra95}. \\
In the alternative Fefferman-Graham formulation where the ambient metric is traded by a Poincaré-Einstein metric in one dimension lower, the conformal structures are realized on the conformal boundary at infinity.
This latter approach, that provides geometric roots for the celebrated AdS/CFT correspondence in physics~\cite{Malda,GKP98,Wit98}, leads to a description of GJMS operators as residues of the scattering operator (aka two-point correlation function in CFT phraseology) as established by Graham and Zworski~\cite{GZ03}.
The (critical) Q-curvature also arises in this context in connection with the volume asymptotics of the Poincaré-Einstein metric. When the dimensionality of the conformal boundary is odd, the renormalized volume is related to the bulk integral of the Q-curvature via the Chern-Gauss-Bonnet formula~\cite{A01, Albin:2005qka,Chang:2005ska}; when the dimensionality of the conformal boundary is even, in turn, the boundary integral of the Q-curvature  is the volume anomaly or, equivalently, the renormalized volume is the conformal primitive of the Q-curvature~\cite{GZ03,HS98,Gra99}.

Now, it was in the study of functional determinants of conformally invariant differential operators, such as the GJMS operators, where the Q-curvature made its first appearance~\cite{BO91}. The infinitesimal variation of the determinant under a conformal (or Weyl) rescaling of the metric reveals the conformal (or Weyl or trace) anomaly; whereas the corresponding finite variation, i.e. its conformal primitive, leads to generalized Polyakov formulas~\cite{Pol81}. The Q-curvature arose in this context as a particular combination of local curvature invariants with a linear transformation law under conformal rescaling of the metric, playing the analog role of the Gaussian curvature on surfaces. Graham~\cite{Gra99} already noticed that the conformal invariance properties of the renormalized volume of a Poincaré-Einstein metric are reminiscent of those for the functional determinants of conformally invariant differential operators, e.g. conformal Laplacian and higher-order GJMS operators, being conformal invariant in odd dimensions but having an anomaly in even dimensions and, on the other hand, those for the volume anomaly are similar to those for the constant term in the expansion of the integrated heat kernel for the conformally invariant differential operator, which vanishes in odd dimensions but in even dimensions is a conformal invariant obtained by integrating a local expression in curvature, namely the conformal anomaly.\\
Remarkably, a `holographic formula' stemming from AdS/CFT heuristics\footnote{The AdS/CFT correspondence certainly predicted the matching of the volume anomaly with the combined conformal anomalies for the free scalars, spinors, and 1-form that enter the four-dimensional vector multiplet of $\mathcal{N}=4$ $SU(N)$ supersymmetric Yang-Mills theory at leading large $N$, as confirmed in~\cite{HS98,HS98-1}. But this connection is somewhat indirect, it relies on non-renormalization theorems of the supersymmetric boundary CFT. In fact, in  six dimensions the matching for the free superconformal $\mathcal{N}=(2,0)$ tensor multiplet is only achieved for the type-B content~\cite{Deser:1993yx} of the Q-curvature, the type-A central charge $a$ is not protected by the supersymmetry so that the combined anomalies do not add up to reproduce the Q-curvature.} provided a direct link between the renormalized volume of the (d+1)-dimensional bulk Poincaré-Einstein metric and functional determinants on the d-dimensional conformal boundary
\begin{equation}
\frac{\det_{-}[-\nabla^2+m^2]}{\det_{+}[-\nabla^2+m^2]}\bigg{|}_{bulk}
=\det\,\langle O_{\lambda} O_{\lambda}\rangle\bigg{|}_{bndry}~
\end{equation}\\
The bulk side contains the one-loop effective action for a massive scalar computed with the resolvent and spectral parameter $\lambda_+=\frac{d}{2}+\nu$ and its analytic continuation to $\lambda_-=\frac{d}{2}-\nu$. The boundary counterpart contains the functional determinant of the two-point function of the dual boundary operator $O_{\lambda}$, a nonlocal integral kernel corresponding to the scattering operator for the radial propagation in the bulk interior. The relation between bulk mass of the scalar field and boundary scaling dimension is, according to the AdS/CFT dictionary, given by $m^2=-\frac{d^2}{4}+\nu^2$.
The formula originated in an attempt to compute an $O(1)$ correction to the partition function under the renormalization group (RG) flow triggered by a boundary double-trace deformation~\cite{Gubser:2002zh,Gubser:2002vv,Hartman:2006dy,Diaz:2007an}. The residues of the scattering operator at its poles become conformally invariant differential operators that in the case of the bulk massive scalar field\footnote{Further extensions of the holographic formula to fields other than the scalar and to quotients of AdS have been studied ever since~\cite{Diaz:2008iv}-\cite{Barvinsky:2017qvf}.} ($\nu\rightarrow k$, $k=1,2,3,...$) correspond to the family of GJMS operators $P_{2k}$
\begin{equation}
\frac{\det_{-}[-\nabla^2-\frac{d^2}{4}+k^2]}{\det_{+}[-\nabla^2-\frac{d^2}{4}+k^2]}\bigg{|}_{bulk}
=\det\, P_{2k}\bigg{|}_{bndry}~
\end{equation}\\
In the conformal class of round metrics on the spheres, the similarities noticed before get promoted to a full-fledged equality because
on the bulk side the volume of Euclidean AdS (or hyperbolic space) factorizes in the effective action due to its homogeneity\footnote{Quotients of AdS, like thermal AdS for example, allow explicit results in terms of Patterson-Selberg zeta functions. In odd dimensions, these examples were also reported in the conformal geometry literature~\cite{Guillarmou05}.}. In this way, for even d, a Polyakov formula for the determinant of the GJMS operators was `holographically' obtained~\cite{Diaz:2008hy} and, perhaps more importantly, the two chief roles of the Q-curvature were directly connected. In particular, a compact formula for the type-A Weyl anomaly coefficient was obtained\footnote{This holographically derived formula for the central charge $a$ was verified later on by using the more standard zeta function regularization combined with Branson's factorization of GJMS operators on the round spheres~\cite{Dowker:2010qy}.} from the bulk Green's function (or resolvent) at coincident points.

A subsequent extension of this clean entry of the AdS/CFT dictionary beyond conformal flatness has remained stalled ever since. Two main obstacles become readily apparent. One is the absence of a viable holographic route to compute the type-B Weyl anomaly in higher-derivative gravities; this is to be contrasted with the simple prescription of evaluating the bulk action at the AdS background to obtain the type-A Weyl anomaly~\cite{ISTY99}.
Second, powers of the Weyl tensor and its derivatives will appear in the heat kernel coefficients to all orders; this is again to be contrasted with the well-known WKB-exactness of the heat kernel in the AdS background~\cite{Camporesi90,Grigorian98,Gopakumar:2011qs} that leaves only the first few terms after resummation.

It is the aim of this note to show how these difficulties can be overcome and to present a holographic derivation of both type-A and type-B Weyl anomaly coefficients for the whole family of GJMS operators in four and six dimensions. We start in Section 2 by first going to a generic compact Einstein manifold on the boundary, exploiting the factorization of GJMS operators into Laplacians, and computing the constant term of their heat kernel expansion in four and six dimensions so as to have the Weyl anomaly beforehand. Section 3 is devoted to the main contribution of this paper, namely the holographic derivation of the Weyl anomaly by considering the heat kernel of the bulk scalar in the corresponding bulk Poincaré-Einstein metric and the resummation that must occur in order to meet the (by now expected) central charges. In the conclusion, Section 4, we summarize and discuss our results. In Appendix A we provide more details about the WKB-exactness and the resummation properties of the bulk scalar heat kernel on the relevant Poincaré-Einstein metric.
\qquad\\

%----------------------------------------------------------------------------------------------------------------------------
\section{Weyl anomaly for GJMS: take I}
\qquad\\
Let us start by examining the GJMS operators on an even d-dimensional compact manifold where the very existence of the ``supercritical'' ones, i.e. $P_{2k}$ with $k>d/2$, is not granted in general. Even if they exist, as in the case of Einstein manifolds, their higher-derivative nature precludes the use of standard heat kernel methods. In the conformal class of round spheres, nevertheless, Branson's factorization of GJMS operators into product of Laplacians~\cite{Bra95} comes to rescue and the type-A Weyl anomaly coefficient can be worked out either by adding the constant terms of the heat expansion for the individual Laplacians or by zeta function regularization~\cite{Dowker:2010qy}.\\
In going beyond the conformally flat class of round metrics on the spheres, as required to access the type-B Weyl anomaly, the leap forward we need is facilitated by Gover's remarkable extension of the factorization of GJMS operators to the more general case of Einstein manifolds~\cite{Gover06}\\

\begin{flalign}%equation}\fl
\setlength\fboxsep{0.3cm}
\setlength\fboxrule{0.5pt}
\boxed{
P_{2k}=\prod_{i=0}^{k-1}\left[-\nabla^2 + \frac{(d+2i)(d-2i-2)}{4d(d-1)}R\right]
}
\end{flalign}%equation}
\vspace{3mm}\\
starting ($i=0$) with the conformal Laplacian or Yamabe operator
\begin{equation}
Y=-\nabla^2 + \frac{d-2}{4(d-1)}R
\end{equation}
The contribution of each Laplacian to the functional determinant, and to the anomaly, can then be computed with standard heat kernel techniques. In addition, as it has already been noticed and successfully put into use ~\cite{Bugini:2016nvn,Beccaria:2017dmw,Acevedo:2017vkk}, although the Einstein condition brings in many simplifications, the curvature invariants that enter the type-B Weyl anomaly remain independent and their coefficients can be efficiently obtained by this shortcut route.

\qquad\\

%------------------------------------------------------------------------
\subsection{Factorization and heat kernel at 4D: two birds, one stone}
\qquad\\
As explained before, a direct way to work out the Weyl anomaly for the GJMS operators is to exploit their factorization on a generic compact Einstein manifold, look for the relevant heat kernel coefficient for each individual factor and then add them all. We will need then the $b_4$ heat coefficient for each of the ``shifted Laplacians'' in the product

\begin{equation}
P_{2k}=\prod_{i=0}^{k-1}\left[-\nabla^2+\frac{(2+i)(1-i)}{12}R\right]
\end{equation}\\
Each shifted Laplacian has the form $-\nabla^2-E$, where $E$ is an endomorphism ~(see e.g. \cite{BFT00} for details) and it is straightforward to get the heat coefficient restricted to the Einstein metric

\begin{equation}
b_4^{(i)}=\left(\frac{i^2(i+1)^2}{288}-\frac{1}{2160}\right)R^2+\frac{1}{180}W^2
\end{equation}\\
Now we simply have to add up the contributions of the individual Laplacians to get the Weyl anomaly for the 4D GJMS operators

\begin{equation}
\label{boundary-4D}
\mathcal{A}_4[P_{2k}]=\sum_{i=0}^{k-1}b_{4}^{(i)}=\left(\frac{k^5}{240}-\frac{k^3}{144}\right)\frac{R ^2}{6}+\frac{k}{180}W^2
\end{equation}\\
Then, regarding the Weyl anomaly basis in 4D, one can trade the Euler density $E_4$ by the Q-curvature ${\mathcal Q}_4$ (type-A) and maintain the Weyl tensor squared $W^2\equiv W_{abcd}W^{abcd}$ which is the obvious independent Weyl-invariant local curvature combination (type-B). The full information on $a$ and $c$ can be gained at one go~\footnote{This is a slightly more efficient way than the usual trick (see, e.g.~\cite{Tseytlin:2013jya}) that restricts first to the round sphere for computing $a$ and then to a Ricci-flat manifold for computing $c-a$.} by considering the generic Einstein metric $g_{_E}$ , since then the Q-curvature reduces to a multiple of the Ricci scalar squared, ${\mathcal Q}_4=R^2/24$, and the Weyl tensor-squared remains unchanged; therefore we have the following rewriting~\cite{Bugini:2016nvn}

\begin{eqnarray}
\mathcal{A}_4 &=&-a\,E_4 \,+\,c\,W^2\\\nonumber
 \\
\nonumber
&=&-4 a\,{\mathcal Q}_4 \,+\,(c-a)\,W^2 \nonumber\\\nonumber
\\
\nonumber
&=&- a\,R^2/6 \,+\,(c-a)\,W^2 \nonumber
\end{eqnarray}\\
Comparing the above relation with the accumulated heat coefficient of the ``shifted Laplacians'', we finally obtain the Weyl anomaly coefficients for the whole GJMS family in 4D

\begin{flalign}%equation}\fl
\setlength\fboxsep{0.3cm}
\setlength\fboxrule{0.5pt}
\boxed{
a_{k}=\frac{k^3}{144}-\frac{k^5}{240}
}
\end{flalign}%equation}

\begin{flalign}%equation}\fl
\setlength\fboxsep{0.3cm}
\setlength\fboxrule{0.5pt}
\boxed{
c_{k}-a_{k}=\frac{k}{180}
}
\end{flalign}\\%equation}
\\
Two remarks are worth mentioning here. First, the quintic polynomial $a_k$ follows as well from the generic expression found in~\cite{Diaz:2008hy} and corroborated by explicit zeta regularization in~\cite{Dowker:2010qy}. Second, only the shifted type-B anomaly coefficient turns out to be linear in $k$ and, in consequence,  meets the holographic expectation of~\cite{Mansfield:1999kk,Mansfield:2003gs} on Ricci-flat backgrounds.
\qquad\\

%------------------------------------------------------------------------
\subsection{Factorization and heat kernel at 6D: four birds, one stone}
\qquad\\
In 6D, we follow the same procedure as in 4D. The factorization of the GJMS operators in terms of ``shifted Laplacians'' is now given by

\begin{equation}
P_{2k}=\prod_{i=0}^{k-1}\left(-\nabla^2+\frac{(3+i)(2-i)}{30}R\right)
\end{equation}
The endomorphism term is $E=-\frac{(3+i)(2-i)}{30}R$ and we denote $d_i=\frac{(3+i)(2-i)}{30}$. The relevant heat-kernel coefficient of the individual Laplacians can be worked out (see e.g.~\cite{BFT00}) and the raw result on a 6D Einstein metric, modulo trivial total derivatives, reads~\footnote{For notation and conventions we refer to~\cite{Bugini:2016nvn}.}

\begin{eqnarray}
%\hspace{15mm}
&b^{(i)}_{6}=& -\frac{d_i^3}{6}R ^3+\frac{d_i^2}{12}R ^3-d_i\left(\frac{1}{180}RRiem^2-\frac{1}{180}RRic^2+\frac{1}{72}R^3\right)\\\nonumber
\\
\nonumber
&&+\frac{1}{7!}\left(-3|\nabla Riem|^2+\frac{44}{9}Riem ^3 - \frac{80}{9}Riem'^3-\frac{16}{3}RicRiem^2\right. \\\nonumber
 \\
\nonumber
&&\left.+\frac{14}{3}RRiem^2-\frac{8}{3}RiemRic^2+\frac{8}{9}Ric^3-\frac{14}{3}RRic^2+\frac{35}{9}R^3\right)\\\nonumber
\end{eqnarray}\\

On the Einstein metric there is a lot of simplifications: the Cotton tensor, the Bach tensor and the traceless part of the Ricci tensor all vanish. Nonetheless,  the type-A and the three type-B terms remain independent~\cite{Bugini:2016nvn}.
We keep a generic 6D Einstein boundary metric $g_{_E}$ so that the Einstein condition reduces the Q-curvature to a multiple of the Ricci scalar cubed,
${\mathcal Q}_6=R^3/225$; the two cubic contractions of the Weyl tensor, denoted by $I_1=W'^{\,3}$ and $I_2=W^3$, remain unchanged; while the third Weyl invariant reduces to
$I_3=W\nabla^2W - \frac{8}{15} R\,W^2$ modulo the trivial total derivative $\frac{3}{2}\nabla^2W^2$ (see e.g.~\cite{Osborn:2015rna}) that we omit in what follows. The 6D Weyl anomaly can then be casted in the following convenient form

\begin{eqnarray}
{\mathcal A}_6&=&-{a}\,E_6\,+\,{c_1}\,I_1\,+\,{c_2}\,I_2\,+\,{c_3}\,I_3 \\
\nonumber\\\nonumber
\qquad\quad&=&-48\,a\,{\mathcal Q}_6+(c_1-96a)I_1+(c_2-24a)I_2+(c_3+8a)I_3\\
\nonumber\\\nonumber
\qquad\quad&=&-16\,a\,R^3/75+(c_1-96a)I_1+(c_2-24a)I_2+(c_3+8a)I_3
\end{eqnarray}

\[%begin{equation*}\fl
\begin{array}{|r c| c| c| c| c| c|} \hline
& \mbox{Curvature invariant } & {\mathcal Q}_6=R^3/225 & I_1 & I_2 &  I_3 \\
\hline {A}_{10}\quad\vline & {R}^{\,3}   &225 &  -&- &- \\
\hline {A}_{11}\quad\vline & {R}{R}ic^{\,2} & 75/2  & -&- &- \\
\hline {A}_{12}\quad\vline & {R}{R}iem^{\,2} &15  &20  &-5 &-5 \\
\hline {A}_{13}\quad\vline & {R}ic^{\,3} & 25/4& -& -& -\\
\hline {A}_{14}\quad\vline & {R}iem \, {R}ic^{\,2} &25/4 & -&- &-  \\
\hline {A}_{15}\quad\vline & {R}ic \, {R}iem^{\,2} & 5/2 &10/3  &-5/6 &-5/6 \\
\hline {A}_{16}\quad\vline & {R}iem^{\,3} & 1 & 4 & 0& -1 \\
\hline {A}_{17}\quad\vline & -{R}iem'^{\,3} &1 & -2& 1/4&1/4  \\
\hline {A}_{5}\quad\vline & |{\nabla}{R}iem|^{2}   &- &-32/3 &8/3 &5/3  \\
\hline
\end{array}
\]%end{equation*}
\\
Making use of the table above to go to the standard anomaly basis and adding up the heat coefficients of the individual Laplacians (tedious but straightforward) we end up with
\begin{eqnarray}
\label{boundary-6D}
\qquad
 7! \; {\mathcal A}_6[P_{2k}]&=&7! \sum_{i=0}^{k-1} b_6^{(i)}\\
\nonumber\\
 &=& -\frac{16}{75} \left(\frac{-3k^7+21k^5-28k^3}{144}\right)R^3\nonumber\\\nonumber\\
 &&+\frac{14(k^3-k)}{9}\left(4I_{1}-I_{2}-I_{3}\right)- \frac{k}{9}\left(24I_{1}-30I_2-13I_{3}\right)\nonumber
\end{eqnarray}\\
From this expression for the accumulated heat coefficients for the shifted Laplacians we finally read off the 6D Weyl anomaly for the whole GJMS tower

\begin{flalign}%equation}\fl
\setlength\fboxsep{0.3cm}
\setlength\fboxrule{0.5pt}
\boxed{
7!a_{k}=-\frac{3k^7-21k^5+28k^3}{144}
}
\end{flalign}%equation}

\begin{flalign}%equation}\fl
\setlength\fboxsep{0.3cm}
\setlength\fboxrule{0.5pt}
\boxed{
7!(c_{1,k}-96a_{k})=\frac{8}{9}k(7k^2-10)
}
\end{flalign}%equation}

\begin{flalign}%equation}\fl
\setlength\fboxsep{0.3cm}
\setlength\fboxrule{0.5pt}
\boxed{
7!(c_{2,k}-24a_{k})=-\frac{2}{9}k(7k^2-22)
}
\end{flalign}%equation}

\begin{flalign}%equation}\fl
\setlength\fboxsep{0.3cm}
\setlength\fboxrule{0.5pt}
\boxed{
7!(c_{3,k}+8a_{k})=-\frac{1}{9}k(14k^2-27)
}
\end{flalign}\\%equation}
\\
Again, two remarks are in order here. First, the polynomial $a_k$ follows also from the generic expression found in~\cite{Diaz:2008hy,Dowker:2010qy}. Second, on Ricci-flat backgrounds the Q-curvature vanishes and $I_3=4\,I_1-I_2$ so that the combined coefficients in front of the two independent Weyl invariant, say $I_1$ and $I_2$, turn out to be linear in $k$,  as can be readily verified,   and therefore agree with the holographic expectation of~\cite{Mansfield:2003bg} (see, also,~\cite{Liu:2017ruz}).
\qquad\\

%----------------------------------------------------------------------------------------------------------------------------
\section{Weyl anomaly for GJMS: take II}
\qquad\\
Let us now turn to our main thrust and try to elucidate the way in which the information on the Weyl anomaly is encoded in the ``hologram'', namely the bulk massive scalar. We proceed in two steps. First, we consider the holographic formula for a bulk Poincaré-Einstein metric with the Einstein metric of before on the boundary conformal class, following the prescription put forward in~\cite{Bugini:2016nvn} that allows to read off the Weyl anomaly coefficient in higher-curvature gravities.

\begin{equation}
\nonumber
 \hat{g}_{_{PE}}=\frac{dx^2+(1-\lambda x^2)^2g_{_E}}{x^2}
\end{equation}
with  $\lambda=\frac{R}{4d(d-1)}$ proportional to the boundary Ricci scalar.\\
At first sight this seems to be of little help because the heat kernel coefficients, in particular those depending on the nonvanishing Weyl tensor, will be present to all orders so that there will be infinitely many higher-curvature terms in the bulk one-loop effective action.\\
In a second step, and despite the above caveat, we compute the Weyl content of the first few heat coefficients. With this partial information at hand and under the crucial assumption of WKB-exactness after resummation, we are able to correctly reproduce the Weyl anomaly coefficients for the whole tower of GJMS in four and in six dimensions, as explained in what follows.
\qquad\\

%------------------------------------------------------------------------
\subsection{Holographic derivation from 5 to 4 dims}
\qquad\\
We consider therefore the holographic formula in the above Poincaré-Einstein metric on the bulk and the corresponding generic compact Einstein metric on the boundary
\begin{equation}
\frac{Z^{^{(-)}}_{_{\text{MS}}}}{Z^{^{(+)}}_{_{\text{MS}}}}\bigg{|}_{_{PE}}\,=\,Z_{_{\text{GJMS}}}\bigg{|}_{E}
\end{equation}
with the bulk one-loop effective action given by the functional determinants of the massive scalar field\footnote{From now on we denote bulk quantities with a hat to distinguish from the corresponding boundary ones.}
 \begin{align}
Z^{^{(+)}}_{_{\text{MS}}}\bigg{|}_{_{PE}}\,=\,\left[ \det\left\{-\hat{\nabla}^{2}+m_k^2\right\}\right]^{-1/2}
\end{align}\\
We first recall the WKB-exact heat expansion in $AdS_5$~\cite{Camporesi90,Grigorian98,Gopakumar:2011qs}. Although there are infinitely many heat kernel coefficients, after factorization of the exponential factor $e^{-4t}$ only the first two remain in five dimensions

\begin{align}
\mbox{massive scalar $m_k^2=k^2-4$: \qquad tr}\,e^{\{\hat{\nabla}^{2}-k^2 +4\}t}\bigg{|}_{_{AdS_5}}\,=\, \frac{1+\frac{2}{3}t }{(4\pi t)^{5/2}}~e^{-k^2t}
\end{align}\\
We need now to depart from $AdS_5$ and determine the pure-Weyl content of the heat kernel on the Poincaré-Einstein metric.
The first contribution arises with $\hat{b}_4$

\begin{align}
\hat{b}_4&\sim \frac{1}{180} \,\hat{W}^2
\end{align}\\
The relevant terms in the next heat coefficient $\hat{b}_6$ are the following

\begin{eqnarray}
%\hspace{15mm}
&\hat{b}_6\sim &\frac{1}{7!}\left(-3|\hat{\nabla}\hat{R}iem|^2+\frac{44}{9}\hat{R}iem ^3 - \frac{80}{9}\hat{R}iem'^3-\frac{16}{3}\hat{R}ic\hat{R}iem^2\right. \\\nonumber
 \\
\nonumber
&&\left.+\frac{14}{3}\hat{R}\hat{R}iem^2-\frac{8}{3}\hat{R}iem\hat{R}ic^2+\frac{8}{9}\hat{R}ic^3-\frac{14}{3}\hat{R}\hat{R}ic^2+\frac{35}{9}\hat{R}^3\right)
\end{eqnarray}\\
We now follow the prescription of~\cite{Bugini:2016nvn} and go to the particular basis of Weyl invariants given by two independent cubic contractions, $\hat{W}^3$ and $\hat{W}'^3$, and the third one given by the 5D Fefferman-Graham invariant $\hat{\Phi}_5=|\nabla \hat{W}|^2-8\hat{W}^2$

\begin{eqnarray}
%\hspace{15mm}
&\hat{b}_6\sim & -\frac{1}{45} \,\hat{W}^2 - \frac{1}{7!}\left(\, \frac{80}{9}\,\hat{W}'^3 -\, \frac{44}{9}\,\hat{W}^3+\,3\,\hat{\Phi}_5\right)
\end{eqnarray}\\
We tabulate the dictionary below for convenience\footnote{The merit of our special basis of curvature invariants is to unveil the direct relation between bulk and boundary Weyl invariants, but of course the contribution of each term of the A-basis has been worked out by other routes in the literature, see e.g.~\cite{Kulaxizi:2009pz,Miao:2013nfa,Beccaria:2015ypa} and references therein.}.

\[%begin{equation*}\fl
\begin{array}{|r c|  c| c| c| c|} \hline
& \mbox{Curvature invariant }  & \hat{\mathit{W}}^2 & \hat{W}'^{\,3}& \hat{W}^{3} &  \hat{\Phi}_5 \\
\hline \widehat{A}_{10}\quad\vline & \widehat{R}^{\,3}   & -& -&- &- \\
\hline \widehat{A}_{11}\quad\vline & \widehat{R}\widehat{R}ic^{\,2}   & -& -&- &- \\
\hline \widehat{A}_{12}\quad\vline & \widehat{R}\widehat{R}iem^{\,2} & -20 & -& -& -\\
\hline \widehat{A}_{13}\quad\vline & \widehat{R}ic^{\,3} & -& -& -& -\\
\hline \widehat{A}_{14}\quad\vline & \widehat{R}iem \, \widehat{R}ic^{\,2} & -&- &- &- \\
\hline \widehat{A}_{15}\quad\vline & \widehat{R}ic \, \widehat{R}iem^{\,2}  & -4 & -& -& -\\
\hline \widehat{A}_{16}\quad\vline & \widehat{R}iem^{\,3} & -6 & -&1 &- \\
\hline \widehat{A}_{17}\quad\vline & -\widehat{R}iem'^{\,3} & 3/2& -1 & - &- \\
\hline \widehat{A}_{5}\quad\vline & |\hat{\nabla}\widehat{R}iem|^{2} &8 &- &- &1 \\
\hline
\end{array}
\]%end{equation*}
\\

After the dust has settled, we realize then that the $-1/45 \hat{W}^2$ in $\hat{b}_6$ can absorbed by the $e^{-4t}$ factor that makes the resummation of the pure-Ricci terms and results in the well-known WKB-exactness of the heat kernel expansion in odd-dimensional hyperbolic space. The remaining Weyl invariant terms in $\hat{b}_6$ do not contribute to the holographic anomaly. Assuming that this WKB-exactness extends to the $\hat{W}^2$ term, the contribution of the one-loop effective Lagrangian of the massive bulk scalar to the holographic Weyl anomaly comes exclusively from the following combination of pure-Ricci (numbers since we set the radius of the asymptotic hyperbolic metric to unity) and pure-Weyl pieces\\
\begin{align}
\int_{0}^{\infty}\frac{dt}{t^{7/2}}e^{-k^2t}\left\{1 + \frac{2}{3}t + \frac{1}{180}\hat{W}^2 t^2 + ...\right\}
\end{align}\\
where the ellipsis stands for higher curvature pure-Weyl invariants that do not contribute to the 4D holographic Weyl anomaly.
After proper time integration we obtain for the one-loop effective Lagrangian (modulo an overall normalization factor that can be easily worked out)\\
\begin{align}
\mathcal{L}^{^{(\text{GJMS})}}_{\text{1-loop}}=\,& \frac{4}{3}\left(\frac{k^5}{5}-\frac{k^3}{3}\right)\cdot\hat{1} + \frac{k}{180}\cdot\hat{W}^2 + ...
\end{align}
The holographic recipe~\cite{Bugini:2016nvn} tells us then how to read the anomaly: the volume part (pure-Ricci) $\hat{1}$ `descends' to the 4D Q-curvature and the pure-Weyl quadratic contraction of the 5D Weyl tensor  `descends' to the analog contraction of the 4D Weyl tensor. In all, the holographic Weyl anomaly one reads off is simply given by
\begin{align}
{\mathcal A}_4[P_{2k}]=&  -4 \left(\frac{k^3}{144}-\frac{k^5}{240}\right)\,{\mathcal Q}_4 + \frac{k}{180}\,W^2
\end{align}\\
in perfect and remarkable agreement with the boundary computation (eqn.\ref{boundary-4D}).
\qquad\\

%------------------------------------------------------------------------
\subsection{Holographic derivation from 7 to 6.}
\qquad\\
We move on now to seven dimensions. The WKB-exact heat expansion in $AdS_7$~\cite{Camporesi90,Grigorian98,Gopakumar:2011qs} requires factorization of the exponential factor $e^{-9t}$ so that only the first three terms remain in seven dimensions

\begin{align}
\mbox{massive scalar $m_k^2=k^2-9$: \qquad tr}\,e^{\{\hat{\nabla}^{2}-k^2 +9\}t}\bigg{|}_{_{AdS_7}}\,=\, \frac{1+2t+\frac{16}{15}t^2 }{(4\pi t)^{5/2}}~e^{-k^2t}
\end{align}\\
To depart from $AdS_7$ and the conformally flat class of bulk and boundary metrics, we consider the pure-Weyl content of the heat kernel on the bulk Poincaré-Einstein metric. The first nontrivial contribution arises again with $\hat{b}_4$

\begin{align}
\hat{b}_4&\sim \frac{1}{180} \,\hat{W}^2
\end{align}
The next contribution comes form the next heat coefficient $\hat{b}_6$

\begin{eqnarray}
%\hspace{15mm}
&\hat{b}_6\sim &\frac{1}{7!}\left(-3|\hat{\nabla}\hat{R}iem|^2+\frac{44}{9}\hat{R}iem ^3 - \frac{80}{9}\hat{R}iem'^3-\frac{16}{3}\hat{R}ic\hat{R}iem^2\right. \\\nonumber
 \\
\nonumber
&&\left.+\frac{14}{3}\hat{R}\hat{R}iem^2-\frac{8}{3}\hat{R}iem\hat{R}ic^2+\frac{8}{9}\hat{R}ic^3-\frac{14}{3}\hat{R}\hat{R}ic^2+\frac{35}{9}\hat{R}^3\right)
\end{eqnarray}\\
The heat coefficients for the scalar Laplacian are universal in the sense that the number in front of each curvature invariant is independent of the dimensionality of the manifold. However, when following the prescription of~\cite{Bugini:2016nvn} and going to the particular basis of Weyl invariants (see table below) given by two independent cubic contractions, $\hat{W}^3$ and $\hat{W}'^3$, and the third one given now by the 7D Fefferman-Graham invariant $\hat{\Phi}_7=|\nabla \hat{W}|^2-8\hat{W}^2$, we obtain a different result

\begin{eqnarray}
%\hspace{15mm}
&\hat{b}_6\sim &\frac{1}{7!}\left(-\, \frac{1916}{9}\,\hat{W}'^3 +\, \frac{503}{9}\,\hat{W}^3-\, 54\,\hat{\Phi}_7\right)
\end{eqnarray}\\

\[%begin{equation*}\fl
\begin{array}{|r c| c| c| c| c|} \hline
& \mbox{Curvature invariant } & \hat{W}'^{\,3} & \hat{W}^{3} &  \hat{\Phi}_7 \\
\hline \widehat{A}_{10}\quad\vline & \widehat{R}^{\,3}   &  -&- &- \\
\hline \widehat{A}_{11}\quad\vline & \widehat{R}\widehat{R}ic^{\,2}  & -&- &- \\
\hline \widehat{A}_{12}\quad\vline & \widehat{R}\widehat{R}iem^{\,2}  & -42 & 21/2& -21/2\\
\hline \widehat{A}_{13}\quad\vline & \widehat{R}ic^{\,3} & -& -& -\\
\hline \widehat{A}_{14}\quad\vline & \widehat{R}iem \, \widehat{R}ic^{\,2} & -&- &-  \\
\hline \widehat{A}_{15}\quad\vline & \widehat{R}ic \, \widehat{R}iem^{\,2} & -6 & 3/2& -3/2\\
\hline \widehat{A}_{16}\quad\vline & \widehat{R}iem^{\,3} & -6 & 5/2&-3/2  \\
\hline \widehat{A}_{17}\quad\vline & -\widehat{R}iem'^{\,3} & 1/2 & -3/8 & 3/8  \\
\hline \widehat{A}_{5}\quad\vline & |\hat{\nabla}\widehat{R}iem|^{2} &8 &-2 &3  \\
\hline
\end{array}
\]%end{equation*}
\\

We now assume WKB-exactness after factorization of the $e^{-9t}$ factor. The convolution with the exponential must absorb  a $-1/20 \hat{W}^2$ contribution to $\hat{b}_6$, that in the 7D case can be rewritten in the Weyl basis $\left[\hat{W}'^3,\hat{W}^3,\hat{\Phi}_7\right]$. In fact, modulo a trivial total derivative $\hat{W}^2=\hat{W}'^3-\frac{1}{4}\hat{W}^3+\frac{1}{4}\hat{\Phi}_7$ on the Poincaré-Einstein metric. So that we obtain, under the assumption of WKB-exactness, the following one-loop effective Lagrangian

\begin{align}
\int_{0}^{\infty}\frac{dt}{t^{9/2}}e^{-k^2t}\left\{1 + 2t + \frac{16}{15}t^2 + \frac{1}{180}\hat{W}^2 t^2 \right.\\
\nonumber\\
\nonumber
\left.+\frac{1}{7!}\left(\frac{352}{9}\,\hat{W}'^3 -\, \frac{64}{9}\,\hat{W}^3 +\,9\,\hat{\Phi}_7\right)t^3+ ...\right\}
\end{align}\\
where again the ellipsis stands for higher-curvature terms in the Weyl tensor that do not contribute to the 6D holographic Weyl anomaly. After proper time integration we obtain for the one-loop effective Lagrangian (modulo an overall normalization factor)

\begin{align}
\mathcal{L}^{^{(\text{GJMS})}}_{\text{1-loop}}=\,& \frac{8}{315} \left(-3k^7+21k^5-28k^3\right)\cdot\hat{1}  \\
\nonumber\\
\nonumber
&-\frac{14k^3}{3\cdot 7!}\cdot\left(4\,\hat{W}'^3-\,\hat{W}^3+\,\hat{\Phi}_7\right)+\frac{k}{9\cdot7!}\cdot\left(352\hat{W}'^3-64\hat{W}^3+81\hat{\Phi}_7\right)+...
\end{align}\\
Now, according to the holographic recipe~\cite{Bugini:2016nvn}, the holographic Weyl anomaly one reads off from the bulk effective Lagrangian is simply

\begin{align}
7! \; {\mathcal A}_6[P_{2k}]=&  -48\,\frac{-3k^7+21k^5-28k^3}{144}\,{\mathcal Q}_6 \\
\nonumber\\
\nonumber
&-\frac{14k^3}{3}\left(4I_{1}-I_{2}+\Phi_{6}\right)+\frac{k}{9}\left(352I_{1}-64I_2+81\Phi_{6}\right)
\end{align}\\
We finally go to the standard basis of 6D Weyl invariants $\left[I_1, I_2, I_3\right]$ by use of the dictionary $3\Phi_{6}=I_3-16I_1+4I_2$

\begin{eqnarray}
\qquad
 7! \; {\mathcal A}_6[P_{2k}]&=& - 48\,\frac{-3k^7+21k^5-28k^3}{144}\,{\mathcal Q}_6\nonumber\\\nonumber\\
 &&+\frac{14k^3}{9}\left(4I_{1}-I_{2}-I_{3}\right)- \frac{k}{9}\left(80I_1-44I_2-27I_{3}\right)\nonumber
\end{eqnarray}\\
and get perfect agreement with the outcome of the boundary computation (eqn.~\ref{boundary-6D}).
\qquad\\
\\
\\
%----------------------------------------------------------------------------------------------------------------------------
\section{Conclusion}
\qquad\\

We have shown the way one bulk Laplacian rules the whole family of boundary GJMS operators and, in particular, the way the conformal anomaly is encoded in the bulk heat kernel. Clearly, the alleged WKB-exactness of the bulk scalar heat kernel on the Poincaré-Einstein metric deserves further analysis and an independent confirmation thereof would be desirable. The boundary computation of the anomaly was facilitated by the factorization of the GJMS operator on a generic Einstein manifold and by the fact that the Einstein condition, besides the many simplifications, does not spoil the independence of the curvature invariants that enter the type-A and type-B Weyl anomaly.

It would be interesting to explore the connection between the one-loop information encoded in the present holographic formula and one-loop Witten diagrams (see e.g.~\cite{Giombi:2017hpr}). For example, one- and two-point correlators of the boundary stress tensor computed from graphs with one and two graviton legs, respectively, with the bulk scalar running in the loop ought to render the $a$ and the $c_T$ central charges\footnote{The coefficient of the two-point function of the stress tensor $c_T$ in 4D is proportional to the $c$ central charge and in 6D,  to $c_3$. In 6D one would need additional (three-point) correlators to disentangle the remaining ($c_1$ and $c_2$) type-B Weyl anomaly coefficients.}.

One subtle feature of the preset computation that we leave as a future direction to look into consists in the following. There is an ambiguity in the construction of GJMS operators given by the addition of terms containing the Weyl tensor. For example, one can add to the Paneitz $P_4$ operator a constant times $W^2$ without changing its conformal properties. In the case of $P_6$ in 6D, besides any of the three Weyl invariants $I_1$, $I_2$ and $I_3$, there is also the freedom to add another term quadratic in the Weyl tensor and in covariant derivatives
(see e.g.~\cite{Osborn:2015rna,Rajagopal:2015lpa}). These additional Weyl terms will certainly modify the conformal anomaly of the differential operators.
The choice  implied by the factorization on Einstein manifolds that we have made use of clearly distinguishes pure-Ricci GJMS with no additional term containing the Weyl tensor. In remains then to be elucidated the way in which the possible additional Weyl terms find their way into the holographic picture.
\qquad\\
\qquad\\
%-----------------------------------------------------------------------------
%\ack
\section*{Acknowledgement}
\qquad\\

We are grateful to S.Acevedo, R.Aros, H.Dorn, S.Dowker, R.Olea, A.Torrielli and A.Tseytlin for valuable discussions.
The work of F.B. was partially funded by grant CONICYT-PCHA/Doctorado Nacional/2014-21140283.
D.E.D. acknowledges support from project UNAB DI 14-18/REG and is also grateful to the Galileo Galilei Institute for Theoretical Physics (GGI) for the hospitality and INFN for partial support during the stay at the program ``New Developments in AdS3/CFT2 Holography'' and to the Quantum Field and String Theory Group at Humboldt University of Berlin for the kind invitation and the opportunity to present the results reported here.
%------------------------------------------------------------------------------------
\appendix
\section{WKB-exactness of the scalar Laplacian}\label{app.A}
\qquad\\
In this appendix, we explicitly compute the first heat coefficients and illustrate the way they get rearranged after factorization of the exponential factor.\\
\paragraph{\bf 5D PE/E}

\begin{align}
\mbox{tr}\,e^{\{\hat{\nabla}^{2}\}t}\bigg{|}_{_{PE}}\,=&\, \frac{1}{(4\pi t)^{5/2}}\left\{ \,1\,-\,\frac{10}{3} \,t\,
+\,\frac{16}{3} \,t^2\,+\,\frac{1}{180} \,\hat{W}^2\,t^2 \right.
\\
\nonumber
\\
\nonumber &\left. -\,\frac{16}{3} \,t^3\,-\,\frac{1}{45} \,\hat{W}^2\,t^3\,- \frac{1}{7!}\left(\, \frac{80}{9}\,\hat{W}'^3 -\, \frac{44}{9}\,\hat{W}^3+\,3\,\hat{\Phi}_5\right)\,t^3\,+\mathcal{O}(t^4)\,\right\}\\
\nonumber
\\
\nonumber
\,=&\, \frac{e^{-4t}}{(4\pi t)^{5/2}}\left\{ \,1\,+\,\frac{2}{3} \,t\, +\,\frac{1}{180} \,\hat{W}^2\,t^2\, \right.
\\
\nonumber
\\
\nonumber &\left. -
\frac{1}{7!}\left(\, \frac{80}{9}\,\hat{W}'^3 -\, \frac{44}{9}\,\hat{W}^3+\,3\,\hat{\Phi}_5\right)\,t^3\,+\mathcal{O}(t^4)\,\right\}
\end{align}\\

\paragraph{\bf 7D PE/E}

\begin{align}
\mbox{tr}\,e^{\{\hat{\nabla}^{2}\}t}\bigg{|}_{_{PE}}\,=&\, \frac{1}{(4\pi t)^{7/2}}\left\{ \,1\,-\,7\,t\,
+\,\frac{707}{30}\,t^2\,+\,\frac{1}{180}\,\hat{W}^2\,t^2 \right.
\\
\nonumber
\\
\nonumber &\left. -\,\frac{501}{10}\,t^3\,-\,\frac{1}{7!}\left(\, \frac{1916}{9}\,\hat{W}'^3 -\, \frac{503}{9}\,\hat{W}^3+\,54\,\hat{\Phi}_7\right)\,t^3\,+\mathcal{O}(t^4)\,\right\}\\
\nonumber
\\
\nonumber
\\
\nonumber
\,=&\, \frac{1}{(4\pi t)^{7/2}}\left\{ \,1\,-\,7\,t\,
+\,\frac{707}{30}\,t^2\,+\,\frac{1}{180}\,\hat{W}^2\,t^2 \right.
\\
\nonumber
\\
\nonumber &\left. -\,\frac{501}{10}\,t^3\,-\,\frac{1}{20}\,\hat{W}^2\,t^3\,+ \frac{1}{7!}\left(\, \frac{352}{9}\,\hat{W}'^3 -\, \frac{64}{9}\,\hat{W}^3+\,9\,\hat{\Phi}_7\right)\,t^3\,+\mathcal{O}(t^4)\,\right\}\\
\nonumber
\\
\nonumber
\\
\nonumber
\,=&\, \frac{e^{-9t}}{(4\pi t)^{7/2}}\left\{ \,1\,+\,2\,t\,+\,\frac{16}{15}\,t^2\, +\,\frac{1}{180} \,\hat{W}^2\,t^2\, \right.
\\
\nonumber
\\
\nonumber &\left. +
\frac{1}{7!}\left(\, \frac{352}{9}\,\hat{W}'^3 -\, \frac{64}{9}\,\hat{W}^3+\,9\,\hat{\Phi}_7\right)\,t^3\,+\mathcal{O}(t^4)\,\right\}
\end{align}\\

\qquad
%------------------------------------------------------------------------------------
%\hspace{0.5cm}
\providecommand{\href}[2]{#2}\begingroup\raggedright\endgroup


\begin{thebibliography}{10}

\bibitem{GJMS92}
C.~R.~Graham, R.~Jenne, L.~J.~Mason and G.~A.~J.~Sparling,
{\it Conformally invariant powers of the Laplacian, I: Existence}, J.\ Lond.\ Math.\ Soc.\ {\bf 46}(1992), 557.

\bibitem{FG85}
C.~Fefferman and C.~R.~Graham, {\it Conformal invariants}, in {\it The Mathematical Heritage
of \'Elie Cartan (Lyon, 1984)}, Ast\'erisque, 1985, Numero Hors Serie, 95-116.

\bibitem{Juhl11}
A.~Juhl,
``Explicit formulas for GJMS-operators and Q-curvatures,''
Geom.\ Funct.\ Analysis {\bf 23} (2013) No. 4, 1278-1370
 [arXiv:1108.0273[math.DG]].

\bibitem{FG13}
C.~Fefferman and C.~R.~Graham,
``Juhl's formulae for GJMS operators and Q-curvatures,''
 J.\ Amer.\ Math.\ Soc. {\bf 26} (2013), 1191-1207
[arXiv:1203.0360[math.DG]].

\bibitem{Bra93}
T.~Branson,
``The Functional Determinant,'' Global\ Analysis\
Research\ Center\ Lecture\ Note\ Series, Number 4, Seoul\ National\
University (1993)

\bibitem{Bra95}
T.~Branson,
``Sharp inequalities, the functional determinant, and
the complementary series,'' Trans.\ Amer.\ Math.\ Soc. {\bf 347}
(1995) 3671.

\bibitem{Malda} J.~M.~Maldacena,
``The large N limit of superconformal field theories and
supergravity,'' Adv.\ Theor.\ Math.\ Phys.\  {\bf 2}, 231 (1998)
[Int.\ J.\ Theor.\ Phys.\  {\bf 38}, 1113 (1999)]
[arXiv:hep-th/9711200];

\bibitem{GKP98}
S.~S.~Gubser, I.~R.~Klebanov and A.~M.~Polyakov, ``Gauge theory
correlators from non-critical string theory,'' Phys.\ Lett.\ B {\bf
428}, 105 (1998) [arXiv:hep-th/9802109];

\bibitem{Wit98}
E.~Witten, ``Anti-de Sitter space and holography,'' Adv.\ Theor.\
Math.\ Phys.\  {\bf 2} (1998) 253 [arXiv:hep-th/9802150].

\bibitem{GZ03}
C.~R.~Graham and M.~Zworski, ``Scattering matrix in conformal
geometry,'' Invent.\ Math. {\bf 152} (2003) 89
[arXiv:math-DG/0109089].

\bibitem{A01}
M.~T.~Anderson, ``$L^2$ curvature and volume renormalization of AHE metrics on 4-manifolds,''
Math.\ Res.\ Lett. {\bf 8} (2001) no. 1-2, 171-188.

\bibitem{Albin:2005qka}
P.~Albin, ``Renormalizing Curvature Integrals on Poincare-Einstein Manifolds,''
  Adv.\ Math.\  {\bf 221} (2009) no.1,  140
 %doi:10.1016/j.aim.2008.12.002
  [math/0504161 [math.DG]].

\bibitem{Chang:2005ska}
  A.~Chang, J.~Qing and P.~Yang,
  ``On the renormalized volumes for conformally compact Einstein manifolds,''
  J.\ Math.\ Sci.\  {\bf 149} (2008) 1755
  [math/0512376 [math.DG]].

\bibitem{HS98}
M.~Henningson and K.~Skenderis, ``The holographic Weyl anomaly,''
JHEP {\bf 9807} (1998) 023 [arXiv:hep-th/9806087].

\bibitem{HS98-1}
M.~Henningson and K.~Skenderis,
``Holography and the Weyl anomaly,''
Fortsch.\ Phys.\ {\bf 48}
(2000) 125 [arXiv:hep-th/9812032].

\bibitem{Deser:1993yx}
S.~Deser and A.~Schwimmer,
``Geometric classification of conformal anomalies in arbitrary dimensions,''
Phys.\ Lett.\ B {\bf 309}, 279 (1993)
[hep-th/9302047].

\bibitem{Gra99}
C.~R.~Graham, ``Volume and area renormalizations for conformally
compact Einstein metrics,'' Rend.\ Circ.\ Mat.\ Palermo (2) Suppl.
No. 63 (2000) 31 [arXiv:math.DG/9909042].

\bibitem{BO91}
T.~Branson and B.~Oersted,
``Explicit functional determinants in four dimensions,''
Proc.\ Amer.\ Math.\ Soc.\ {\bf 113} (1991) 669.

\bibitem{Pol81}
 A.~M.~Polyakov,
 ``Quantum Geometry of Bosonic Strings,''
Phys.\ Lett.\ B {\bf 103} (1981) 207 [Phys.\ Lett.\  {\bf 103B} (1981) 207].
%doi:10.1016/0370-2693(81)90743-7.

\bibitem{Gubser:2002zh}
S.~S.~Gubser and I.~Mitra,
``Double trace operators and one loop vacuum energy in AdS / CFT,''
Phys.\ Rev.\ D {\bf 67} (2003) 064018
% doi:10.1103/PhysRevD.67.064018
[hep-th/0210093].

\bibitem{Gubser:2002vv}
S.~S.~Gubser and I.~R.~Klebanov,
``A Universal result on central charges in the presence of double trace deformations,''
Nucl.\ Phys.\ B {\bf 656} (2003) 23
% doi:10.1016/S0550-3213(03)00056-7
[hep-th/0212138].

\bibitem{Hartman:2006dy}
T.~Hartman and L.~Rastelli,
``Double-trace deformations, mixed boundary conditions and functional determinants in AdS/CFT,''
JHEP {\bf 0801} (2008) 019
% doi:10.1088/1126-6708/2008/01/019
[hep-th/0602106].

\bibitem{Diaz:2007an}
D.~E.~Diaz and H.~Dorn,
``Partition functions and double-trace deformations in AdS/CFT,''
JHEP {\bf 0705} (2007) 046
% doi:10.1088/1126-6708/2007/05/046
[hep-th/0702163 [HEP-TH]].

\bibitem{Diaz:2008iv}
D.~E.~Diaz,
``Holographic formula for the determinant of the scattering operator in thermal AdS,''
J.\ Phys.\ A {\bf 42} (2009) 365401
% doi:10.1088/1751-8113/42/36/365401
[arXiv:0812.2158 [hep-th]].

\bibitem{Aros:2009pg}
R.~Aros and D.~E.~Diaz,
``Functional determinants, generalized BTZ geometries and Selberg zeta function,''
J.\ Phys.\ A {\bf 43} (2010) 205402
% doi:10.1088/1751-8113/43/20/205402
[arXiv:0910.0029 [gr-qc]].

\bibitem{Aros:2011iz}
R.~Aros and D.~E.~Diaz,
``Determinant and Weyl anomaly of Dirac operator: a holographic derivation,''
J.\ Phys.\ A {\bf 45} (2012) 125401
% doi:10.1088/1751-8113/45/12/125401
[arXiv:1111.1463 [math-ph]].

\bibitem{Dowker:2013mba}
J.~S.~Dowker,
``Spherical Dirac GJMS operator determinants,''
J.\ Phys.\ A {\bf 48} (2015) no.2,  025401
% doi:10.1088/1751-8113/48/2/025401
[arXiv:1310.5563 [hep-th]].

\bibitem{Giombi:2013yva}
S.~Giombi, I.~R.~Klebanov, S.~S.~Pufu, B.~R.~Safdi and G.~Tarnopolsky,
``AdS Description of Induced Higher-Spin Gauge Theory,''
JHEP {\bf 1310} (2013) 016
% doi:10.1007/JHEP10(2013)016
[arXiv:1306.5242 [hep-th]].

\bibitem{Giombi:2013fka}
S.~Giombi and I.~R.~Klebanov,
``One Loop Tests of Higher Spin AdS/CFT,''
JHEP {\bf 1312} (2013) 068
% doi:10.1007/JHEP12(2013)068
[arXiv:1308.2337 [hep-th]].

\bibitem{Tseytlin:2013jya}
A.~A.~Tseytlin,
``On partition function and Weyl anomaly of conformal higher spin fields,''
Nucl.\ Phys.\ B {\bf 877}, 598 (2013)
[arXiv:1309.0785 [hep-th]].

\bibitem{Tseytlin:2013fca}
A.~A.~Tseytlin,
``Weyl anomaly of conformal higher spins on six-sphere,''
Nucl.\ Phys.\ B {\bf 877}, 632 (2013)
[arXiv:1310.1795 [hep-th]].

\bibitem{Giombi:2014iua}
S.~Giombi, I.~R.~Klebanov and B.~R.~Safdi,
``Higher Spin AdS$_{d+1}$/CFT$_d$ at One Loop,''
Phys.\ Rev.\ D {\bf 89} (2014) no.8,  084004
% doi:10.1103/PhysRevD.89.084004
[arXiv:1401.0825 [hep-th]].

\bibitem{Beccaria:2014jxa}
M.~Beccaria, X.~Bekaert and A.~A.~Tseytlin,
``Partition function of free conformal higher spin theory,''
JHEP {\bf 1408} (2014) 113
% doi:10.1007/JHEP08(2014)113
[arXiv:1406.3542 [hep-th]].

\bibitem{Aros:2014xga}
R.~Aros, F.~Bugini and D.~E.~Diaz,
``On Renyi entropy for free conformal fields: holographic and q-analog recipes,''
J.\ Phys.\ A {\bf 48} (2015) 105401
% doi:10.1088/1751-8113/48/10/105401
[arXiv:1408.1931 [hep-th]].

\bibitem{Beccaria:2014xda}
M.~Beccaria and A.~A.~Tseytlin,
``Higher spins in AdS$_{5}$ at one loop: vacuum energy, boundary conformal anomalies and AdS/CFT,''
JHEP {\bf 1411} (2014) 114
% doi:10.1007/JHEP11(2014)114
[arXiv:1410.3273 [hep-th]].

\bibitem{Beccaria:2014qea}
M.~Beccaria, G.~Macorini and A.~A.~Tseytlin,
``Supergravity one-loop corrections on AdS$_7$ and AdS$_3$, higher spins and AdS/CFT,''
Nucl.\ Phys.\ B {\bf 892} (2015) 211
% doi:10.1016/j.nuclphysb.2015.01.014
[arXiv:1412.0489 [hep-th]].

\bibitem{Beccaria:2015vaa}
M.~Beccaria and A.~A.~Tseytlin,
``On higher spin partition functions,''
J.\ Phys.\ A {\bf 48} (2015) no.27,  275401
% doi:10.1088/1751-8113/48/27/275401
[arXiv:1503.08143 [hep-th]].

\bibitem{Beccaria:2015uta}
M.~Beccaria and A.~A.~Tseytlin,
``Conformal a-anomaly of some non-unitary 6d superconformal theories,''
JHEP {\bf 1509} (2015) 017
% doi:10.1007/JHEP09(2015)017
[arXiv:1506.08727 [hep-th]].

\bibitem{Barvinsky:2005ms}
A.~O.~Barvinsky and D.~V.~Nesterov,
``Quantum effective action in spacetimes with branes and boundaries,''
Phys.\ Rev.\ D {\bf 73} (2006) 066012
% doi:10.1103/PhysRevD.73.066012
[hep-th/0512291].

\bibitem{Barvinsky:2014kta}
A.~O.~Barvinsky,
``Holography beyond conformal invariance and AdS isometry?,''
J.\ Exp.\ Theor.\ Phys.\  {\bf 120} (2015) no.3,  449
% doi:10.1134/S1063776115030036
[arXiv:1410.6316 [hep-th]].

\bibitem{Barvinsky:2017qvf}
A.~O.~Barvinsky,
``Extended Holography: Double-Trace Deformation and Brane-Induced Gravity Models,''
Russ.\ Phys.\ J.\  {\bf 59} (2017) no.11,  1788.
% doi:10.1007/s11182-017-0977-y

\bibitem{Guillarmou05}
C.~Guillarmou,
``Generalized Krein formula, determinants and Selberg zeta function in even dimension,''
American Journal of Math. 131 (2009), no 5. [Arxiv math.SP/0512173] .

\bibitem{Diaz:2008hy}
D.~E.~Diaz,
``Polyakov formulas for GJMS operators from AdS/CFT,''
JHEP {\bf 0807} (2008) 103
% doi:10.1088/1126-6708/2008/07/103
[arXiv:0803.0571 [hep-th]].

\bibitem{Dowker:2010qy}
J.~S.~Dowker,
``Determinants and conformal anomalies of GJMS operators on spheres,''
J.\ Phys.\ A {\bf 44} (2011) 115402
% doi:10.1088/1751-8113/44/11/115402
[arXiv:1010.0566 [hep-th]].

\bibitem{ISTY99}
C.~Imbimbo, A.~Schwimmer, S.~Theisen and S.~Yankielowicz,
``Diffeomorphisms and holographic anomalies,''
Class.\ Quant.\ Grav.\  {\bf 17} (2000) 1129
[arXiv:hep-th/9910267].

\bibitem{Camporesi90}
R.~Camporesi,
``Harmonic analysis and propagators on homogeneous spaces,''
Phys.\ Rept.\ {\bf 196} (1990) 1.

\bibitem{Grigorian98}
A.~Grigor'yan and M.~Noguchi
``The heat kernel on hyperbolic space,''
Bulletin of LMS, {\bf 30} (1998) 643-650.

\bibitem{Gopakumar:2011qs}
R.~Gopakumar, R.~K.~Gupta and S.~Lal,
``The Heat Kernel on $AdS$,''
JHEP {\bf 1111} (2011) 010
% doi:10.1007/JHEP11(2011)010
[arXiv:1103.3627 [hep-th]].

\bibitem{Gover06}
A.~R.~Gover
`` Laplacian Operators and Q-curvature on Conformally Einstein Manifolds,''
Math. Ann. (2006) 336: 311
https://doi.org/10.1007/s00208-006-0004-z
[arXiv:math/0506037 [math.DG]].

\bibitem{Bugini:2016nvn}
F.~Bugini and D.~E.~Diaz,
``Simple recipe for holographic Weyl anomaly,''
JHEP {\bf 1704} (2017) 122
% doi:10.1007/JHEP04(2017)122
[arXiv:1612.00351 [hep-th]].

\bibitem{Beccaria:2017dmw}
M.~Beccaria and A.~A.~Tseytlin,
``C$_{T}$ for higher derivative conformal fields and anomalies of (1, 0) superconformal 6d theories,''
JHEP {\bf 1706} (2017) 002
%doi:10.1007/JHEP06(2017)002
[arXiv:1705.00305 [hep-th]].

\bibitem{Acevedo:2017vkk}
  S.~Acevedo, R.~Aros, F.~Bugini and D.~E.~Díaz,
``On the Weyl anomaly of 4D Conformal Higher Spins: a holographic approach,''
JHEP {\bf 1711} (2017) 082
 %doi:10.1007/JHEP11(2017)082
[arXiv:1710.03779 [hep-th]].

\bibitem{BFT00}
F.~Bastianelli, S.~Frolov and A.~A.~Tseytlin,
``Conformal anomaly of (2,0) tensor multiplet in six dimensions and  AdS/CFT
correspondence,'' JHEP {\bf 0002} (2000) 013
[arXiv:hep-th/0001041].

\bibitem{Tseytlin:2013jya}
A.~A.~Tseytlin,
``On partition function and Weyl anomaly of conformal higher spin fields,''
Nucl.\ Phys.\ B {\bf 877}, 598 (2013)
[arXiv:1309.0785 [hep-th]].

\bibitem{Osborn:2015rna}
H.~Osborn and A.~Stergiou,
``Structures on the Conformal Manifold in Six Dimensional Theories,''
JHEP {\bf 1504} (2015) 157
% doi:10.1007/JHEP04(2015)157
[arXiv:1501.01308 [hep-th]].

\bibitem{Mansfield:1999kk}
P.~Mansfield and D.~Nolland,
``One loop conformal anomalies from AdS / CFT in the Schrodinger representation,''
JHEP {\bf 9907} (1999) 028
% doi:10.1088/1126-6708/1999/07/028
[hep-th/9906054].

\bibitem{Mansfield:2003gs}
P.~Mansfield, D.~Nolland and T.~Ueno,
``The Boundary Weyl anomaly in the N=4 SYM / type IIB supergravity correspondence,''
JHEP {\bf 0401} (2004) 013
% doi:10.1088/1126-6708/2004/01/013
[hep-th/0311021].

\bibitem{Mansfield:2003bg}
P.~Mansfield, D.~Nolland and T.~Ueno,
``Order 1 / N**3 corrections to the conformal anomaly of the (2,0) theory in six-dimensions,''
Phys.\ Lett.\ B {\bf 566} (2003) 157
% doi:10.1016/S0370-2693(03)00777-9
[hep-th/0305015].

\bibitem{Liu:2017ruz}
J.~T.~Liu and B.~McPeak,
``One-Loop Holographic Weyl Anomaly in Six Dimensions,''
JHEP {\bf 1801} (2018) 149
% doi:10.1007/JHEP01(2018)149
[arXiv:1709.02819 [hep-th]].

\bibitem{Kulaxizi:2009pz}
M.~Kulaxizi and A.~Parnachev,
``Supersymmetry Constraints in Holographic Gravities,''
Phys.\ Rev.\ D {\bf 82} (2010) 066001
% doi:10.1103/PhysRevD.82.066001
[arXiv:0912.4244 [hep-th]].

\bibitem{Miao:2013nfa}
R.~X.~Miao,
``A Note on Holographic Weyl Anomaly and Entanglement Entropy,''
Class.\ Quant.\ Grav.\  {\bf 31} (2014) 065009
% doi:10.1088/0264-9381/31/6/065009
[arXiv:1309.0211 [hep-th]].

\bibitem{Beccaria:2015ypa}
M.~Beccaria and A.~A.~Tseytlin,
``Conformal anomaly c-coefficients of superconformal 6d theories,''
JHEP {\bf 1601} (2016) 001
% doi:10.1007/JHEP01(2016)001
[arXiv:1510.02685 [hep-th]].

\bibitem{Giombi:2017hpr}
S.~Giombi, C.~Sleight and M.~Taronna,
``Spinning AdS Loop Diagrams: Two Point Functions,''
JHEP {\bf 1806} (2018) 030
% doi:10.1007/JHEP06(2018)030
[arXiv:1708.08404 [hep-th]].

\bibitem{Osborn:2015rna}
H.~Osborn and A.~Stergiou,
``Structures on the Conformal Manifold in Six Dimensional Theories,''
JHEP {\bf 1504} (2015) 157
% doi:10.1007/JHEP04(2015)157
[arXiv:1501.01308 [hep-th]].

\bibitem{Rajagopal:2015lpa}
S.~Rajagopal, A.~Stergiou and Y.~Zhu,
``Holographic Trace Anomaly and Local Renormalization Group,''
JHEP {\bf 1511} (2015) 216
% doi:10.1007/JHEP11(2015)216
[arXiv:1508.01210 [hep-th]].

\end{thebibliography}
\end{document}